\documentclass[english,twocolumn]{article}
\usepackage[T1]{fontenc}
\usepackage{lmodern}
\usepackage[big]{dgruyter}
\usepackage[utf8]{inputenc}
\usepackage[caption=false]{subfig}
\usepackage{mathrsfs}
\usepackage{accents}
\usepackage[normalem]{ulem}
\usepackage[dvipsnames]{xcolor}

\usepackage[all]{hypcap}
\usepackage{enumerate}
\usepackage{microtype}
\usepackage{float}
\usepackage{graphicx}
\usepackage{amssymb,amsfonts}
\usepackage{amsmath}
\usepackage{textgreek}
\usepackage{placeins}
\usepackage{siunitx}
\DeclareSIUnit[number-unit-product=]\percent{\char`\%} 

\begin{document}

\articletype{Research article}

\author*[1]{Apostolos Apostolakis}
\author[2]{Mauro F. Pereira}
\affil[1]{Department of Condensed Matter Theory, Institute of Physics, Czech Academy of Sciences,  Na Slovance 1999/2, 
182 21 Prague, Czech Republic, email: apostolakis@fzu.cz. https://orcid.org/0000-0002-9080-5405}
\affil[2]{1. Department of Condensed Matter Theory, Institute of Physics, Czech Academy of Sciences,  Na Slovance 1999/2, 
182 21 Prague, Czech Republic. 2.Department of Physics, Khalifa University of Science and Technology, Abu Dhabi 127788, UAE, email: mauro.pereira@ku.ac.ae. https://orcid.org/0000-0002-2276-2095}

\title{Superlattice nonlinearities for Gigahertz-Terahertz generation in harmonic multipliers}
\abstract{Semiconductor superlattices are strongly nonlinear media offering several technological challenges associated with the generation of  high-frequency Gigahertz radiation and very effective frequency multiplication up to several Terahertz. However, charge accumulation, traps and interface defects lead to pronounced asymmetries in the nonlinear current flow, from which high harmonic generation stems. This problem requires a full non-perturbative solution of asymmetric current flow under irradiation, which we deliver in this paper within  the Boltzmann-Bloch approach. We investigate the nonlinear output on both frequency and time domains and demonstrate a significant enhancement of even harmonics by tuning the interface quality. Moreover, we find that increasing arbitrarily the input power is not a solution for high nonlinear output, in contrast with materials described by conventional susceptibilities. There is a complex combination of asymmetry and power values leading to maximum high harmonic generation.  
}


\maketitle

\section{Introduction}

\begin{figure}[t]
\centering
\includegraphics[scale=1.3]{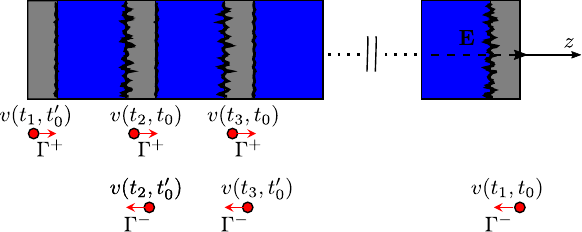}
   \caption{Schematic diagram of the origin of asymmetric current flow in a  semiconductor superlattice due to differences in interface roughness depending on the sequence of layers. GaAs layers are depicted in blue (color online) and AlAs layers in grey (color online). The interface of GaAs over AlAs has lower quality than that of AlAs over GaAs.  The different relaxation rates $\Gamma^+$ ($\Gamma^-$) depend on the direction of the time-dependent miniband velocity $v(k_z(t_i),k_z(t_0))$. Here $t_0$ and $t_0'$ designate just different starting times [see Eq. (\ref{eq:tdcurren})].  The SSL sample is biased by an electric field, $\mathbf{E}=(0,0,E_{ac}\cos(2\pi\nu t))$ parallel  to the direction of the $z$--axis.   } 
\label{fig1}
\end{figure}

The inherent nonlinearities of electronic systems can be exploited for the development of novel compact sources in the terahertz (THz) region \cite{dhillon20172017,tonouchi2007cutting,blanchard2010generation,chen2012nonlinear,vaks2012high}. The very same nonlinearities and their underlying microscopic origin serve as sensitive means for controlling high harmonic generation (HHG) processes. A notable very recent example is the generation of THz harmonics in a single-layer graphene due to hot Dirac fermionic dynamics under low-electric field conditions \cite{hafez2018extremely}. 
In a parallel effort, advances in strong-field and attosecond physics  have paved the way to HHG in bulk crystals operating in a highly nonperturbative regime \cite{schubert2014sub,Langer2016,drescher2001x,smirnova2009high,krausz2009attosecond}.  The first experimental observation of non-perturbative HHG in a bulk crystal was explained on the
basis of a simple two-step model in which the nonlinearity stemmed from the anharmonicity of electronic motion in the band combined with multiple Bragg reflections at the zone boundaries \cite{ghimire2011observation}. The high frequency  (HF) nonlinearities  which contribute to  harmonic upconversion  in bulk semiconductors have been  associated to dynamical Bloch oscillations combined with coherent interband polarization processes \cite{schubert2014sub,hohenleutner2015real}. The aforementioned  models allow the use of tight-binding dispersions \cite{golde2008high}  to describe the electronic band  and therefore the radiation from a nonlinear intraband current. One of the systems that demonstrate a similar highly nonparabolic energy dispersion are man-made semiconductor superlattices (SSLs) \cite{esaki1970superlattice,wacker2002semiconductor}.  In fact, the possibility of spontaneous frequency multiplication due to the effect of  nonparabolicity  in a SSL miniband structure   was first predicted in the early works of Esaki-Tsu \cite{tsu1971nonlinear} and Romanov \cite{romanov1972nonlinear}. Superlattices are created by alternating layers of two semiconductor materials with similar lattice constants resulting in the formation of a spatial periodic potential. Furthermore,  SSLs host rich dynamics in the presence of a driving field, which include the formation of Stark ladders \cite{mendez1988stark},  the manifestation of Bragg reflections and Bloch oscillations \cite{waschke1993coherent}. From the viewpoint of applications,  SSLs have attracted great interest because they allow the development of devices which operate at microwave \cite{khosropanah2009phase} and far-infrared frequencies \cite{khosropanah2009phase,vaks2012high} suitable for high precision spectroscopic studies and  detection of submillimeter waves.  In addition, a considerable number of  studies have tackled the task of engineering  parametric amplifiers \cite{renk2005subterahertz,renk2007semiconductor} and frequency multipliers \cite{klappenberger2004semiconductor,paveliev2012experimental,pereira2017theory} based on superlattice periodic structures. Note that  although  the first semiconductor superlattice frequency multipliers (SSLM) were developed  for the generation of microwave radiation \cite{grenzer1995microwave},   significant progress has been achieved combining high-frequency operation (up to 8.1 THz, $\sim$ 50th harmonic) \cite{paveliev2012experimental,vaks2012high} and high conversion efficiency \cite{endres2007application} comparable to the performance of Schottky diodes \cite{endres2007application,siles2010physics}.
\\There are various mechanisms that  contribute to the HF nonlinearities of SSL devices. Once this distinction has been clearly made, it is simple to connect  the underlying physical mechanisms to the frequency multiplication effects. It was found that spontaneous multiplication takes place in a dc biased tight-binding SSL, when the Bloch-oscillating electron wave packet is driven by the input oscillating field \cite{winnerl2000frequency,schomburg1996superlattice}. Moreover, the increase of optical response \cite{winnerl2000frequency}  was  due to  the frequency modulation of Bloch oscillations \cite{ignatov1995thz,romanov2005bloch} which  arise in the negative differential conductivity (NDC) region of the current-voltage characteristic, i.e. the current decreases with increasing bias.
On the other hand, if a SSL device is in a NDC state, the nonlinearities can be further enhanced by the onset of high-field domains \cite{le1992gunn,schomburg1998current} and the related propagation phenomena \cite{makarov2015sub} in a similar way as the electric-field domains in bulk semiconductors \cite{gunn1964instabilities}. Thus, the ultrafast creation and annihilation of electric domains during the time-period of an oscillating field contributes to harmonic generation processes in SSLs \cite{scheuerer2003frequency}.  This type of dynamics has been found to depend on plasma effects \cite{pamplin1970negative,klappenberger2004ultrafast} induced by the space-charge instabilities and the dielectric relaxation time processes which dictate the exact conditions for the NDC state \cite{klappenberger2004ultrafast}. The expected THz response from Bloch oscillations  in a miniband SSL, under the influence of a THz electric field, might also  deviate due to strong excitonic effects \cite{meier1994coherent,dignam1999excitonic,wang2008tunable}. 
Harvesting the nonlinearities discussed above can potentially lead to more efficient SSLMs or other devices suitable for achieving extremely flexible frequency tuning. \\
Our approach is inspired by very recent theoretical and experimental investigations \cite{pereira2017theory,pereira2017terahertz,apostolakis2019controlling,apostolakis2019devices} of SSLM behavior, which revealed the development of even harmonics due to imperfections in the superlattice structure.  In general, when a adequately strong oscillating field couples energy into the SSLM in the absence of constant bias, only odd harmonics are emitted. However, Ref. \cite{pereira2017theory}  showed that symmetry breaking was induced by asymmetric current flow and scattering processes under forward and reverse bias.  This approach combined nonequilibrium Green's function calculations with an Ansatz solution of the Boltzmann equation in the relaxation rate approximation. Furthermore, the asymmetric scattering rates were attributed to  the  different elastic (interface roughness) scattering  rates  which have risen from  the non-identical  qualities of the SSL interfaces. In general, elastic scattering processes can have a significant effect on  the electron transport in semiconductor superlattices.The conventional method to study  the role of elastic scattering on miniband transport and generation of high-frequency radiation \cite{ignatov1991transient,alekseev1998spontaneous}  are the one dimensional (1D) SSL Balance equations \cite{ignatov1991transient} which can be extended to address two-dimensional and three-dimensional \cite{gerhardts1993effect, romanov2002self}  SSL transport and optical properties. They cannot, however, include systematically the different scattering processes under forward and reverse bias. A handful of experiments have been devoted to examine the harmonics of current oscillations \cite{schomburg1998current}, transient THz response \cite{ferreira2009boltzmann}, dephasing mechanisms of Bloch oscillations \cite{unuma2006dephasing} and the electron mobility \cite{sakaki1987interface} under the influence  of isotropic-elastic-scattering time.
\\In this paper, we elucidate how the effects of  asymmetric scattering processes could be used to control  the implications of the SL potential on the response of miniband electrons to an oscillating electric field $E(t)=E_{ac}\cos(2\pi \nu t)$. We solve the Boltzmann transport equation (BTE) by the method of characteristics \cite{chambers1952kinetic,ignatov1976nonlinear}, which  allows to address explicitily  the asymmetric intraminiband relaxation processes in semiconductor  superlattices. Earlier in Refs.~\cite{shmelev1997current,shmelev1998high}  the relaxation rate was assumed to depend  on the electron velocity allowing to estimate analytically the high-frequency conductivity of an asymmetric superlattice but with resorting  to  perturbative analysis of the Boltzmann equation.  Before proceeding further it is worthwhile first to highlight the main points of this work:\\
(i) We eliminate the numerical instabilities  which originate from the Ansatz solution of Ref.  \cite{pereira2017theory, footnote2}.\\
(ii) Our solution delivers a time-domain analysis of the mechanism responsible for the built up of high harmonic generation.\\
(iii) We theoretically demonstrate that the multiplication effects can be effectively controlled by  special designs of superlattice interfaces (asymmetric elastic scattering). \\  
The anisotropic effects \cite{pereira2017theory} reflect  that  typically the interfaces of a host material (A) grown on an different host material (B) are found to be rougher than those of B on A (see Fig. \ref{fig1}), indicating grading or intermixing of the constituent materials between SSL layers \cite{feenstra1994interface,tokura1992anisotropic}. 
\\ This paper is organised as follows. Section   \ref{sec:level2} provides an overview of a semiclassical theory describing the charge transport in SSLs in the presence of asymmetric scattering. In Sec. \ref{sec:level3}, we discuss the nonlinear optical response  of miniband electrons in an asymmetric SL structure and we present  results of exact numerical simulations describing the spontaneous HHG. Complementary insight is provided next with time-domain calculations. In Appendix  \ref{App0},  we revisit in more detail the   expressions related to the solution of the BTE which is implemented in this work. On the other hand, Appendix \ref{App1} entails analytical solutions for HHG based on an Ansatz that can lead to numerical instabilities to highlight our  more efficient solution.

\section{\label{sec:level2} Semiclassical formulation}

Throughout this work we use the  standard energy dispersion, $\epsilon(k_z)=\epsilon^a-2\mid T \mid \cos (k_z d)$, which describes the kinetic energy carried by an electron in the lowest SL miniband \cite{esaki1970superlattice,wacker2002semiconductor}. Here $\epsilon_a$ is the center of the miniband, $\mid T \mid$  is the miniband quarter-width, $k_z$ is the projection of crystal momentum on the $z$--axis (axis parallel to the general grown direction) and $d$ is the superlattice period. Note that in this transport model, the effects of inter-miniband tunneling are neglected. To simulate the temporal distribution function $f(\textbf{k},t)$ of the single electron, we employ  a  semiclassical approach based on the Boltzmann transport equation \cite{wacker2002semiconductor} 

\begin{equation}
\dfrac{\partial f}{\partial t}+\dfrac{\textbf{F}}{\hbar}\dfrac{\partial f}{\partial  \textbf{k}}=I\left\lbrace f \right\rbrace,
\label{eq:bte1}
\end{equation}
where  $\mathbf{F}$ is the force $(-e)E$ corresponding to a time dependent electric field  in the $z$--direction of the SSL and $\textbf{k}$ is the total momentum which can decomposed into $k_z$ and the quasimomentum in the $x$--$y$ plane $k_\parallel=(k_x,k_y)$. The  right-hand side term of Eq. (\ref{eq:bte1}) represents the collision integral. Instead of using a single relaxation rate approximation model equivalent to $I\left\lbrace f \right\rbrace =-(f-f_0)\Gamma/\hbar$ \cite{Ashcroft} with $f_0 (\mathbf{k})$ being the equilibrium fermi distribution, we will resort to two scattering rates $\Gamma$ to adequately describe the asymmetric relaxation processes. The asymmetric elastic scattering would result to enhanced scattering processes into certain directions.
Thus, the kinetic equation can be rewritten in  the following form
\begin{equation}
\textrm{\L}^{\pm}f=\dfrac{\Gamma^{\pm}f_0}{\hbar}
\label{eq:bte2}
\end{equation}
where $\textrm{\L}^{\pm}= 1/\hbar\left(\textbf{F}\partial/\partial\textbf{k}+\hbar \partial/\partial t+\Gamma^{\pm} \right)$ are  integral operators corresponding  to the different relaxation rates ($\Gamma^+$ and $\Gamma^-$). By using the inverse of the operator, $\textrm{\L}^{-1}$, on the left of Eq. (\ref{eq:bte2}), we obtained the time-dependent current density

\begin{eqnarray}\nonumber
 j(t)=\frac{2e}{(2\pi)^3 }\int d^3k\hspace{0.1cm}{f}_0(\mathbf{k})  \hspace{0.1cm}\int_{-\infty}^{t}dt_0  \hspace{0.1cm}  \hspace{0.1cm}\dfrac{\Gamma (t_0) v(t,t_0)}{ \hbar \Delta(t,t_0)} \\ \hspace{-2 cm} \times \exp\left\lbrace-\int_{t_0}^{t}\dfrac{\Gamma(y)}{\hbar} dy \right\rbrace, \hspace{1 cm}
\label{eq:tdcurren}
\end{eqnarray}
\begin{figure}[t]
\includegraphics[scale=0.3]{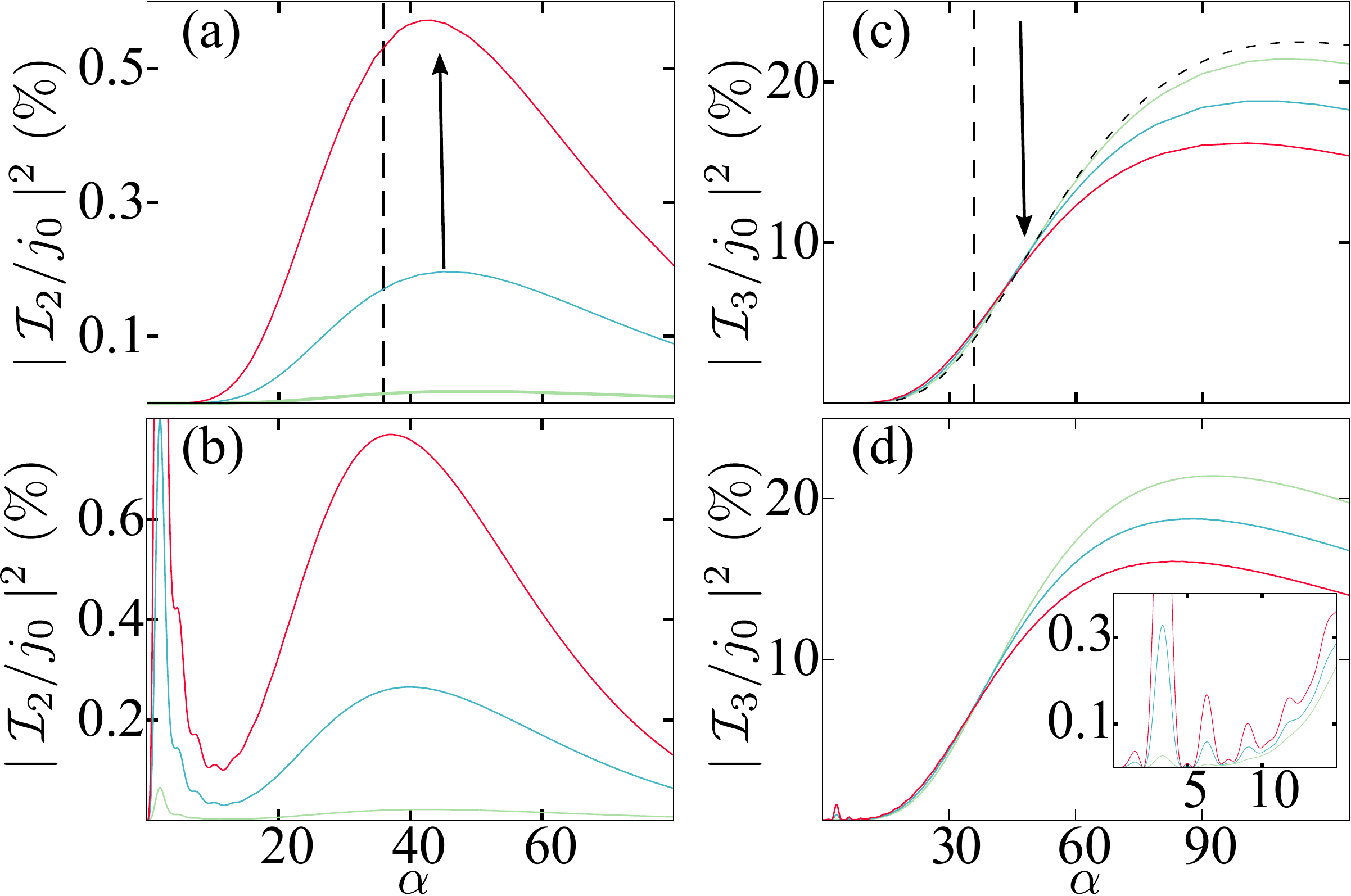}
\centering
   \caption{(Color online) Comparison of the nonlinear response $\mid\mathcal{I}_l(\nu)\mid^2$ characterizing the generation of second [panels (a) and (b)] and third [panels (c)  and (d)] harmonics. The curves are calculated using either the path integral approach of Eq. \ref{eq:tdcurren} [(a) and (c)], or the ansatz of Eq. \ref{eq:ansatz} [(b) and (d)] as a function of the parameter $\alpha=eE_{ac}d/(h \nu)$ and different values of the asymmetry coefficient $\delta=1.05,1.2, 1.4$. The dashed curve (c) indicates the usual third harmonic at $\delta=$1 whereas the vertical dashed lines [(a) and (c)] designate the critical field ($a=a_c$) for which the SSL can operate in the NDC part of the VI characteristic.  The inset zooms on  numerical instabilities for  $\mid\mathcal{I}_3(\nu)\mid^2$ with small parameter $\alpha$ in the third harmonic. In all cases the frequency of the oscillating field is $\nu=$ 141 GHz.} 
\label{fig2}
\end{figure}
where $k_z$ is integrated over the Brillouin zone, the integration limits of the in-plane  components $k_\parallel$ are $\pm \infty$ and $\Delta(t,t_0)$ controls the level of current flow asymmetry.  Equation (\ref{eq:tdcurren}) summarizes that an electron which passes through the point  $\mathbf{k}$ at time $t$, follows different  collisionless trajectories which takes it through the points  $\mathbf{k}(t_0)$ at times $t_0<t$. This compact solution requires the assumption of the following  condition for the relaxation energy
\begin{equation}
     \Gamma = 
    \begin{cases}
     \Gamma^+& \text{$v(t,t_0)>0$,} \\[5pt]
      \Gamma^-& \text{$v(t,t_0)<0$} .
      \label{eq:bconditions}
    \end{cases}       
\end{equation}
Here $v(t,t_0)$ represents the time-dependent miniband velocity  which reveals the propagation direction of the electron along the sample and therefore indicates the interaction  with the high-quality or low-quality interface [see Fig. \ref{fig1}].  For a further discussion of Eqs. (\ref{eq:bte2})-(\ref{eq:bconditions}) see Appendix \ref{App0}.  The time dependence of the velocity $v(t,t_0)$ is obtained from the set of the  equations 
\begin{subequations}
\label{eq:all-fp_tilde}
\begin{eqnarray}
\dfrac{d k_z(t,t_0)}{dt}=\dfrac{2 \pi \nu}{d}  \alpha \cos(2\pi \nu t), \hspace{0.1cm} k_z(t_0,t_0)=k_{z_0}  , \label{eq:fp_xtilde}\\
v(t,t_0)=\dfrac{2\mid T\mid d}{\hbar} \sin k_z d, \hspace{0.65cm} z(t_0,t_0)=0. \label{eq:fp_ptilde}
\end{eqnarray}
\end{subequations}
where the parameter $\alpha=eE_{ac}d/(h\nu)$ critically affects the strength of the nonlinear optical response and consequently  the harmonic-conversion properties of the SSLM. 
\\ The Fourier transform of the time-dependent current [Eq. (\ref{eq:tdcurren})] can be used to obtain spectral peaks at multiples of the driving frequency. In particular, the intensity of the emitted radiation from the SSL structure is determined by the Poynting vector, which is proportional to the harmonic current term \cite{pereira2017theory}
 \begin{eqnarray}
\mathcal{I}^2_l(\nu)=2 \lbrace \langle j(t)\cos(2 \pi \nu l t) \rangle^2 +\langle j(t)\sin(2 \pi \nu l t) \rangle^2 \rbrace,
\label{eq:PoyntVec}
\end{eqnarray}
where the integration $\langle..\rangle$ signifies time-averaging  over time interval of infinite time in the general case. Nevertheless, considering that the current response is induced merely by a monochromatic field $(\nu)$, it is sufficient to average only over the time-period $T_{\nu}=1 / \nu$.  Both even and odd harmonics are present in a biased SSL due to symmetry breaking. However, in this paper we focus on symmetry breaking due structural effects leading to asymmetric current flow. Thus in all numerical results for harmonic generation, there is no static electric field, i.e.  $E_{dc}d=$0 and the Bragg reflections from minizone boundaries are not assigned to Bloch oscillations (BO) with the conventional oscillation frequency $v_B=eE_{dc}d/h$. In contrast, the Bragg scattering is manifested as frequency modulation of electron oscillations during a cycle ($T_{\nu}=1/\nu$) of the oscillating field.   Oscillations of this type  are known as Bloch oscillations in  a harmonic field (BOHF) \cite{romanov2005bloch}. Combining the Bloch acceleration theorem [Εq. (\ref{eq:fp_xtilde})] and the energy dispersion $\epsilon(k_z d)$, one can find  the dependence of miniband energy on time
\begin{eqnarray}
\epsilon(t)=\epsilon_0 -4\mid T \mid \sum_{n=1}^{\infty} J_{2n}(\alpha) \cos \lbrace2n(2\pi \nu t)\rbrace,
\label{eq:timee}
\end{eqnarray}
where $\epsilon_0=\epsilon^a-2 \mid T \mid J_{0}(\alpha)$ and $J_{0}(.)$, $J_{2n}(.)$ are bessel functions of the first kind. From Eq. (\ref{eq:timee}) we see that the electron energy oscillates within the miniband with frequency $\nu^{2n}_{\epsilon}$ at every even mupltiple of $\nu$: $\nu^{2n}_{\epsilon}=ln$ with $l=2, 4, 6, 8...$  On the contrary,   the miniband velocity varies  at every odd multiple of $\nu$: $\nu^{2n+1}_{v}=ln$ with $l=3, 5, 7, 9...$ stemming directly from the equation $v(t)=4\mid T \mid d/\hbar \sum_{n=0}^{\infty} J_{2n+1}(\alpha) \sin \lbrace(2n+1)(2\pi \nu t)\rbrace$ which is derived, just as Eq. (\ref{eq:timee}), from a real-valued expression of Jacobi-Anger expansion
\cite{abramowitz1988handbook}.  Due to the  collisions 
the electron experiences damped BOHF and thus the latter equation can be used as an input for the solution of the BTE [see Eq. (\ref{eq:tdcurren})]. Hereafter, it  becomes clear that the Boltzmann-Bloch transport theory requires  a self-consistent solution of the Boltzmann transport equation and the semiclassical equations of motion obeyed by Bloch momentum $k_z$. In the quasistatic limit the electron performs high quality BOHF when $\alpha>\alpha_c$. Here $\alpha_c=U_c/(h\nu)$ with $U_c=e E_{ac}d=\Gamma$ designating the energy required from the ac-field in order to bring temporarily the SSL to an active state equivalent to the NDC region of the VI characteristic 
result to the formation of high electric field domains which act as additional linearities. We must underline that in this work the high-frequency field considered to be acting on the superlattice leads to a single-electron state and the electric field within the SSL remains uniform. As a result, the gain at some harmonics of the oscillating field is related only to the nonlinearity of the voltage-current characteristic and the BOHF oscillations.
The condition [Eq. (\ref{eq:bconditions})] dictates the different relaxation times $\Gamma(t)=\Gamma^+$ or $\Gamma^-$, reflecting on the coefficient $\Delta(t,t_0)$=1 or $\delta$ in Eq. (\ref{eq:tdcurren}).
This asymmetry coefficient $\delta=\Gamma^+/\Gamma^-$  plays an important role in this work, since it indicates the differences between the interfaces leading to deviation from the perfectly anti-symmetric voltage of the Esaki and Tsu model \cite{esaki1970superlattice}. See Fig. \ref{fig7} per se in Appendix \ref{App1}. 
An increase of  $\delta$ can be interpreted as a structural variation of the initial SSL structure.
In the present work  we assume that $\delta\geq1$ which implies that the flow from left to right will be favored over the flow from right to left.  Furthermore, the asymmetry coefficient depends on the elastic and inelastic scattering rates, which are either determined from measured values \cite{schomburg1998current,patane2002tailoring}  or nonequilibrium Green's functions calculations \cite{pereira2017theory,apostolakis2019controlling}.  It is important to notice that similar kinetic formulas  to Eq. (\ref{eq:tdcurren}) have been used to treat the different types of scattering processes in superlattices \cite{fromhold2004chaotic,greenaway2009controlling}. However, none of these works have systematically included a tensor analyzing the different relaxation processes which correspond to an  asymmetric SSL structure. In this paper, the values of the SSL parameters in Eqs. (\ref{eq:tdcurren})-(\ref{eq:all-fp_tilde}) are taken from recent experiments and predictive simulations \cite{pereira2017theory,pereira2017terahertz}:
 $d=6.23$ nm,  $\mid T \mid =30$ meV, $\Gamma^+$=21 meV, $j_0$=2.14$\times10^9 $ $\mathrm{A}/\mathrm{m}^2$ and  the corresponding critical field is $E_c^+=3.4$ kV/mm.  For the integration of the equation of motions (\ref{eq:fp_xtilde}), (\ref{eq:fp_ptilde})
\begin{figure*}[t]
\centering
    \includegraphics[width=0.85\textwidth]{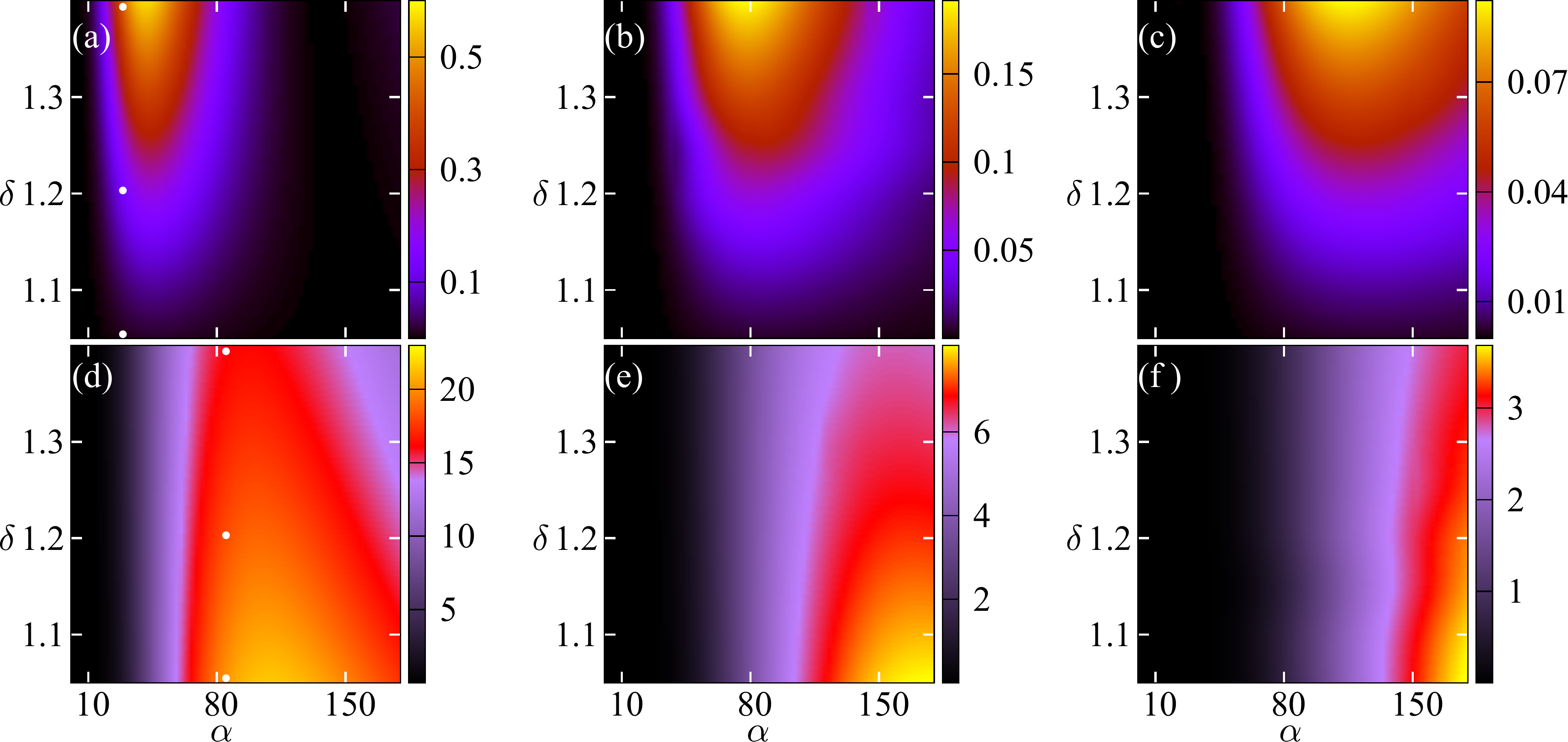}
    \caption{Color maps showing the dependence of the harmonic emission $\mid\mathcal{I}_l(\nu)\mid^2$ on the parameter $\alpha$ and asymmetry coefficient $\delta$ for (a) $l=2$, (b) $l=3$, (c) $l=4$, (d) $l=5$, (e) $l=6$ and (f) $l=7$  harmonics. The white dots in panels (a) and (d) correspond to the values of $\alpha$ and $\delta$ used in the time domain calculations depicted in Fig. \ref{fig6}(c) and Fig. \ref{fig6}(d)  respectively. The color bar is normalized to the peak current $j_0$. The color maps were calculated using Eqs. (\ref{eq:tdcurren})-(\ref{eq:PoyntVec}).} 
\label{fig3}
\end{figure*}
we implement the classical fourth-order Runge-Kutta method known for its stability \cite{press1992numerical}.\\
Before moving forward with results, we should make a brief recap of a previous research.  A NEGF approach, in which the different interfaces were described by using an interface roughness self-energy, gave good agreement with static current voltage, but could not be implemented for a GHz input. Thus, this predictive input was used in a hybrid NEGF-Boltzmann equation approach by employing a analyical Ansatz solution for the asymmetric current flow \cite{pereira2017theory,pereira2017terahertz,apostolakis2019controlling,apostolakis2019devices}. However, the Ansatz leads in some cases to numerical instabilities and errors as shown in Fig. \ref{fig2}. This is one of the main motivations of this paper, which delivers a clean numerical solution that does not need the Ansatz. For correctness the analytical Ansatz solution is described in Appendix \ref{App1}. 
\section{Results \label{sec:level3}}

\begin{figure}[t]
\centering
\includegraphics[scale=0.5]{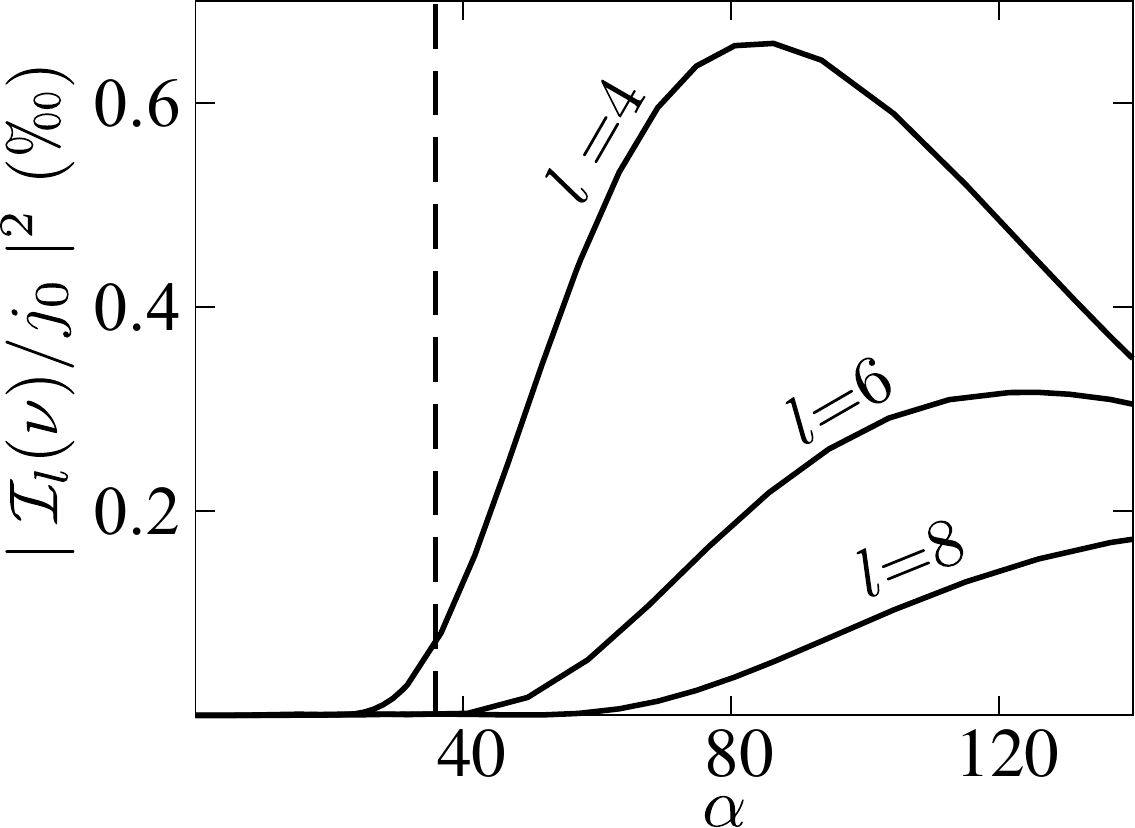}
\caption{High-order even harmonics  as a function of the parameter $\alpha$. The  asymmetry parameter is $\delta=1.2$  and  the frequency of the oscillating field is $\nu=141$ GHz. The dashed curve signifies the critical value $a_c$ for which the SSL can operate in the NDC part of the VI characteristic.} 
\label{fig4}
\end{figure}

\begin{figure}[t]
\centering
\includegraphics[scale=0.6]{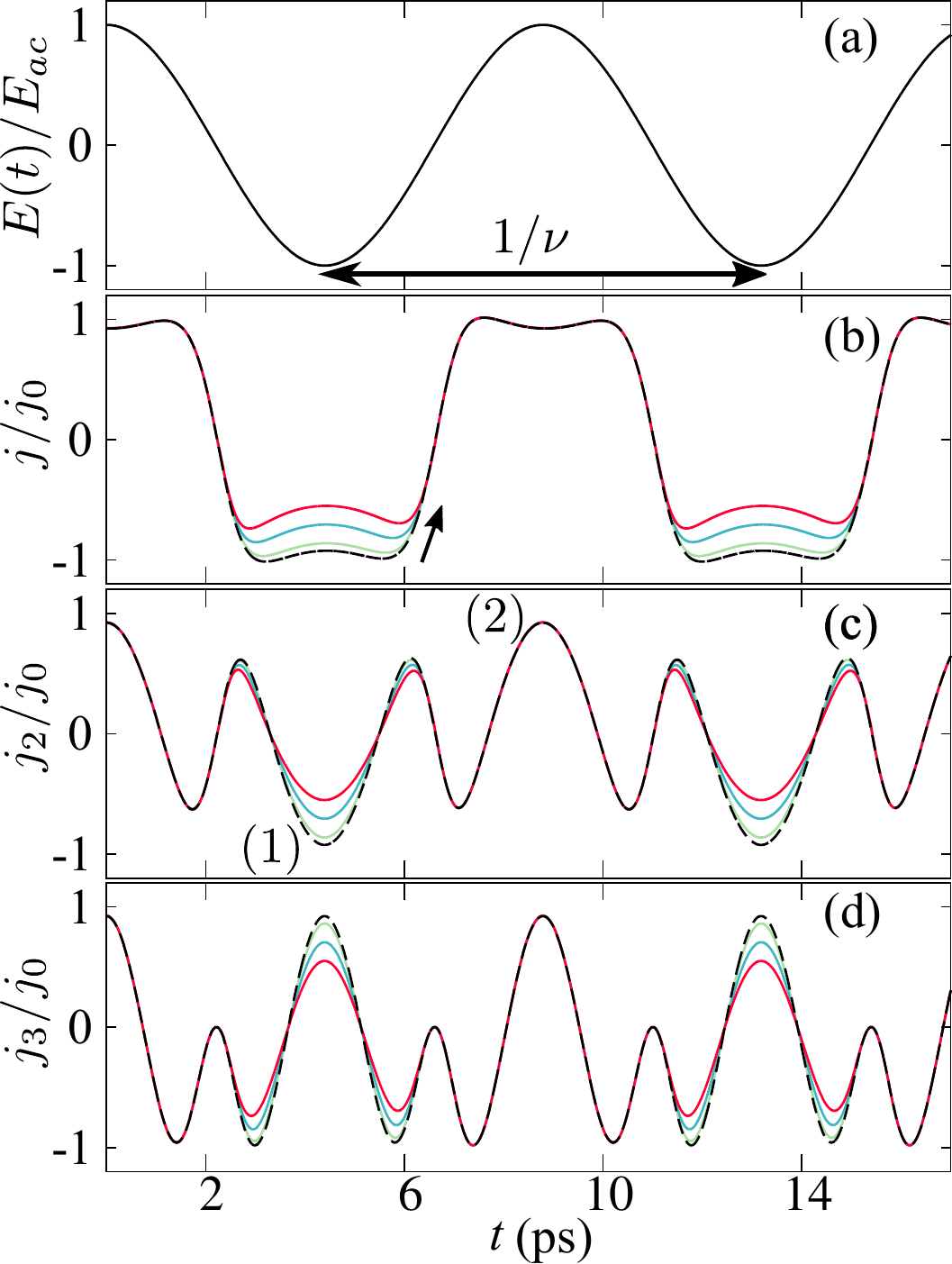}
   \caption{(Color online) Nolinear response of miniband electrons  by considering asymmetric  scattering processes. (a) The normalized electric field [$E(t)/E_{ac}$] which causes the time dependent drift.  (b) The time-dependent current $j(t)$ [see Eq. (\ref{eq:tdcurren})] is depicted over two cycles of the input field $E(t)$. (c) The second-harmonic $j_2(t)$  and (d) the third-harmonic current oscillations $j_3(t)$ calculated for different values of the asymmetry parameter $\delta=1,1.05,1.2, 1.4$. The labels (1) and (2) denote relevant local minima and maxima. In all cases,  the value of the parameter $\alpha \simeq 27 $ corresponds to an electric field with amplitude $E_{ac}$=0.75   $E_{c}$ and oscillating frequency $\nu=141$ GHz. The arrow marks increasing asymmetry. } 
\label{fig5}
\end{figure}

\begin{figure}[t]
\centering
\includegraphics[scale=0.6]{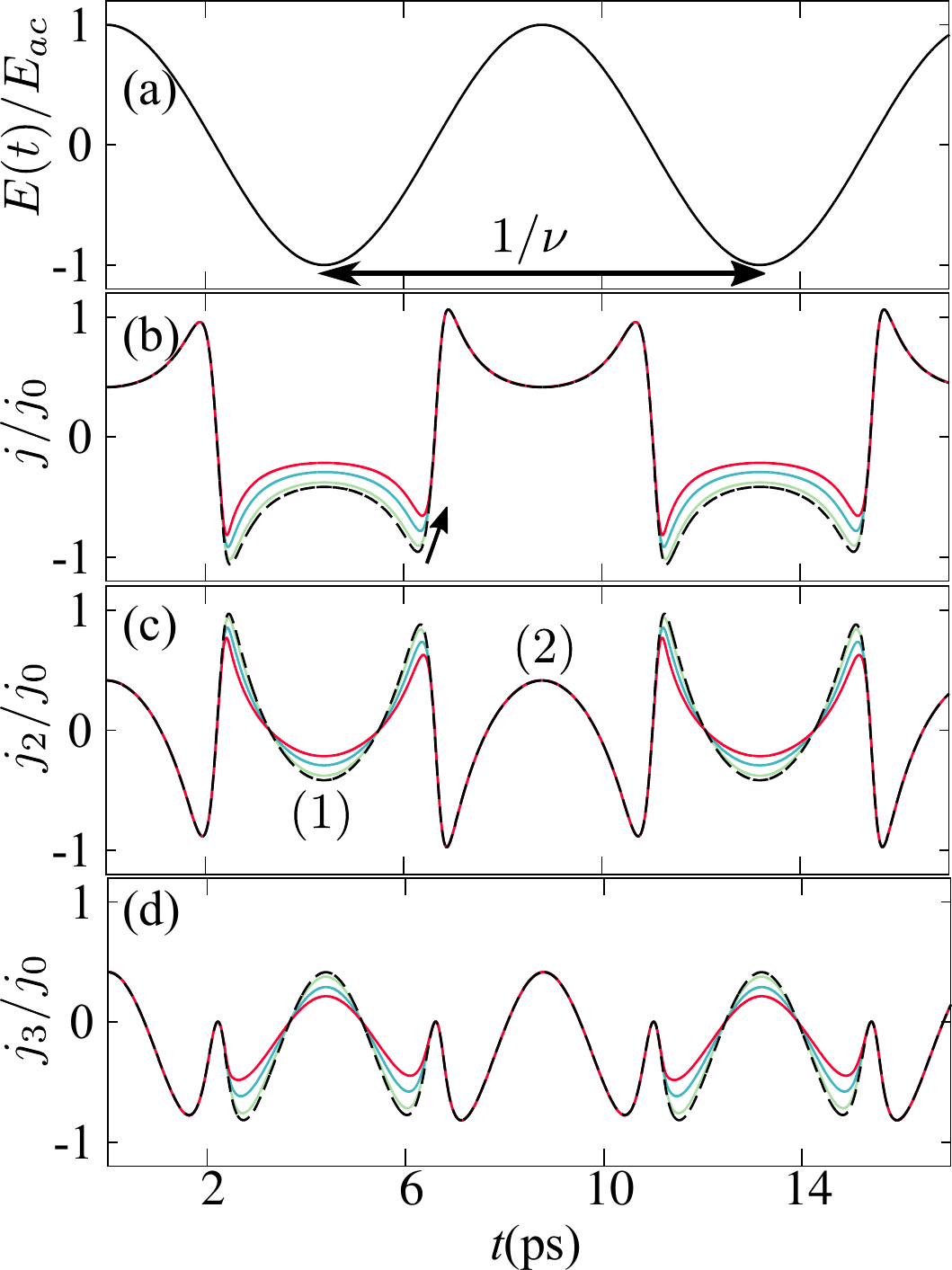} 
   \caption{(Color online) Nolinear response of miniband electrons  by considering asymmetric  scattering processes. (a) The normalized electric field [$E(t)/E_{ac}$] which causes the time dependent drift.  (b) The time-dependent current $j(t)$ [see Eq. (\ref{eq:tdcurren})] is depicted over two cycles of the input field $E(t)$. (c) The second-harmonic $j_2(t)$  and (d) the third-harmonic current oscillations $j_3(t)$ calculated for different values of the asymmetry parameter $\delta=1,1.05,1.2, 1.4$. The labels (1) and (2) denote relevant local minima and maxima. In all cases,  the value of the parameter $\alpha \simeq 86 $ corresponds to an electric field with amplitude $E_{ac}=$ 2.4  $E_{c}$ and oscillating frequency $\nu=141$ GHz. The arrow marks increasing asymmetry. } 
\label{fig6}
\end{figure}

In this section we will investigate the effects of asymmetric scattering on HHG by implementing  the  approach developed in the previous section and compare its predictions to those of the analytical Ansatz solution. The basic idea is to vary the asymmetry coefficient $\delta$ which in both approaches is defined as the ratio of the different relaxation rates ($\Gamma^{\pm}$) and then examine the effects on  even and odd order harmonics.   Figure \ref{fig2} depicts the second harmonic (left-handed panels) and third harmonic output (right-handed panels)  as a function  of the $\alpha$ parameter for different values of the asymmetry coefficient $\delta$. The dependencies $\mid\mathcal{I}_l(\nu)\mid^2(\alpha)$ were calculated using Eqs. (\ref{eq:analcur1})--(\ref{eq:ansatz}) and Eqs.  (\ref{eq:tdcurren})--(\ref{eq:PoyntVec}) in Fig. \ref{fig2}(a) and Fig. \ref{fig2}(b), respectively.   Both approaches yield similar results for the second harmonic in a wide range of $\alpha$. In particular, we highlight that asymmetric relaxation times are an unconventional mechanism for frequency doubling in SSLs. As $\delta$ increases, the frequency doubling effects become more pronounced and, eventually, give rise  to stronger optical response almost up to 0.6 $\%$. We note that the Ansatz solution may, however, contribute to nonphysical  numerical instabilities by revealing intense second harmonic generation even at small amplitudes of the oscillating field.  Therefore, the numerical solution offers a reliable way to treat the scattering induced asymmetries in the current flow. Now we turn our attention to the third harmonic output in the presence  of asymmetric current flow which is quite different from the behavior of the second harmonic output. Moreover, increasing $\delta$ suppresses it, implying an redistribution of spectral components in favor of even harmonics as shown in Fig. \ref{fig2}(d). One can see that the maximum output of the third  harmonic might be potentially reduced from 20  to 15 $\%$. In this case the  different solutions appear to be more consistent with each other. However, as shown in the inset of Fig. \ref{fig2}(d) the Ansatz solution  can lead to  numerical instabilities comparable with the maximum output of the second harmonic. Once we  established that the novel approach developed  in this work affords the significant variation of the asymmetry coefficient, we can have an in-depth look into the HHG processes.\\
Further insight on how asymmetric effects can result in a significant gain at some even harmonic frequencies and suppression at some other odd-order harmonics, is given by the color maps in Fig. \ref{fig3}. It shows the calculated values $\mid\mathcal{I}_l(\nu)\mid^2$  as a function of $\alpha$ and $\delta$. The black area indicates  values ($\alpha$, $\delta$) for which $\mathcal{I}_l$ exhibits small or negligible  harmonic response. The colored areas unfold distinct islands of significant harmonic response. For example, Fig. \ref{fig3}(a) reveals significant enhancement of $I_2$ for   $1.2\lesssim \delta \lesssim 1.4$  and $10\lesssim \alpha \lesssim 70$.  On the other hand,  in the same region Fig. \ref{fig3}(d)  demonstrates a weak third harmonic response of the irradiated superlattice. The corresponding island of enhanced $I_3$ is shifted to significantly  larger $\alpha$ values. The magnitude of $I_3$ increases  approximately from $\delta=1.2$ to $\delta=1$ and thus obtaining a maximum for a SSL structure with  perfectly symmetric interfaces. The width of the islands changes significantly for the higher-order even harmonics as shown in Figs. \ref{fig3}(b) and \ref{fig3}(c). However, although the width of the colored islands is increased, the strength of the harmonic content is reduced, by an order of magnitude for $I_6$ [Fig. \ref{fig3}(c)]  in comparison with $I_2$ [Fig. \ref{fig3}(a)]. The colored areas of the higher-odd harmonics are notably suppressed in the regions where higher-even harmonics are being developed. Therefore, in order to achieve easily detectable  odd harmonics the SSL should operate deep inside the NDC region. At this point it is important to highlight that the higher the harmonic order, the larger the input power must be. However, note that arbitrarily increasing the input power is not a solution for high nonlinear output, in contrast with materials described by conventional susceptibilities. There is a complex combination of asymmetry and power values leading to maximum HHG generation.  For example, Fig. \ref{fig4} demonstrates the output of higher even-order harmonics (beyond the 2nd harmonic) which drastically drops when the input power is significantly larger. The SSL device after excitation by a strong GHz input signal can generate measurable 8th harmonic up to $\sim$ 0.02 $\%$. The magnitude of the emitted power in units  \textmugreek W is related to  harmonic term $\mathcal{I}_l$ as $P_l(\nu)$= $\mathcal{T}$ $\mathcal{I}^2_l(\nu)$ where the coefficient $\mathcal{T}=A \hspace{0.05cm}\mu_0\hspace{0.05cm} c\hspace{0.05cm} L^2/(8 n_r)$ obtained from  the time-averaged Poynting vector by neglecting the waveguide effects. Here $\mu_0$ is the permeability and $c$ is the speed of light by considering both of them in  free-space. For  typical mesa area $A=$(10  $\times$ 10) $\mu  \mathrm{m}^2$, effective path length through the crystal $L=$121.4 nm and refractive index $n_r$=$\sqrt{13}$ (GaAs), one can obtain $\mathcal{T}\simeq 77$ \textmugreek W. Now it is straightforward to calculate the emitted power corresponding to Figs. (\ref{fig2}-\ref{fig4}). As a consequence, for a value $\alpha \simeq 34$ close to but below the  $\alpha_c$ the emitted power can reach the values $P_2=0.4$ \textmugreek W and $P_4(\nu)=0.01$ \textmugreek W  at room temperature for the second  and fourth harmonic respectively. These magnitudes indicate that significant gain can appear at second and fourth-order harmonics in the absence of electric domains which might affect the HHG processes when $\alpha>\alpha_c$. 
\\ Next, we complement the steady-state analysis with calculations of the time-dependent nonlinear response of the miniband electrons. Our time-dependent solution [see Eq. \ref{eq:tdcurren}] can provide further insight in the frequency-conversion of the input signal related to the asymmetric scattering processes.
Figure  \ref{fig5} depicts  the oscillating field,  the nonlinear current oscillations,  the  second harmonic component  and the third harmonic component  which   occur in the presence of asymmetric scattering rates. The oscillating field $E(t)$ [see Fig. \ref{fig5}(a)] causes a time-dependent electron drift with a time dependent current $j(t)$ which contains different harmonic components  due to the enhanced nonlinear response as shown in  Fig. \ref{fig5}(b). 
In a perfectly symmetric structure, the irradiation of the superlattice with input radiation leads only to odd-order multiplication and therefore the second harmonic signal $j_2(t)=j(t)\cos(4 \pi\nu t)$ [dashed curve in Fig. \ref{fig5}(c)] averaged over time is $<j_2>_t=0$. On the contrary, for a higher asymmetry parameter $\delta$ (arrowed), the time realization of $j_2(t)$ demonstrates oscillations whose amplitude is highly asymmetric. In this case, the the first peak (1)  becomes sufficiently smaller than peak (2) resulting in $<j_2>_t$ different than zero as is evident from Fig. \ref{fig5}(c).  The third harmonic component in the current is due to the  BOHF which stem from the anharmonic motion of the electron within the miniband. Every half-period ($T_{\nu}/2$), $j_3$ contributes a phase of an opposite sign  with respect to the temporal evolution of the electric field [see Fig. \ref{fig5}(a), (d)]. With increasing asymmetry coefficient $\delta$, the amplitude of the arrowed peak is reduced, which leads to suppression of the third harmonic component $ <j_3>_t $.\\ For a electric field with sufficiently larger amplitude but with the same oscillating  frequency, the current response becomes evidently more anharmonic [see Fig. \ref{fig6}(b)]. This has important implications for both second and third-order harmonics and serious consequences in the case of increasing the asymmetry parameter $\delta$.

On the one hand, the increase of the asymmetry between the two relaxation rates results in more pronounced differences between the oscillations amplitudes (1), (2) of the second harmonic $j_2(t)$ and their adjacent peaks [Fig. \ref{fig6}(c)]. Consequently, the second harmonic is  suppressed for a larger $E_{ac}$ but still enhanced for a different $\delta$. On the other hand,  the $\alpha$ parameter being larger than $\alpha_c$ would induce higher quality BOHF and therefore larger third harmonic components [Fig. \ref{fig6}(d)]. We note though that a larger  asymmetry will reduce the emission of $j_3$  due to the strong suppression of the  closely neighboring peaks to the main one. \\ Before summarizing the main results of this paper, it is noteworthy to highlight further immediate applications of our approach. Evidently the electron energy   oscillates within the miniband with frequency at every even multiple of the frequency of the oscillating electric field [see Eq. (\ref{eq:tdcurren})]. The behaviour of effective electron mass, which explicitly depends on energy \cite{ignatov1993esaki}, varies significantly in the presence of highly asymmetric scattering and thus the generated    harmonics   could   be linked to the concept of negative effective mass similar to \cite{shorokhov2015physical}. Significant production of even harmonics has been previously predicted for an electrically excited  SSL due to parametric amplification \cite{hyart2006terahertz} or other parametric processes \cite{romanov2000self}  which stem from the existence of an internal electric field in the  structure. In this respect, it is interesting to  study how the parametric processes can affect the harmonic generation  \cite{hyart2007theory} or the Bloch gain \cite{romanov2000self} profile in the presence of asymmetric current flow. Moreover, our approach has a great potential for analyzing the effects of asymmetric scattering processes on the intensity of harmonics by means of externally applied voltages and/or intense ultrafast optical pulses \cite{ferreira2009boltzmann,wang2008tunable}. Finally, the predictions in the present work  highlight the prospects for the systematic study of asymmetric effects in different superlattice systems including coupled superlattices in a synchronous state \cite{matharu2013high}  and, even more generally, in other multilayer structures such as high temperature superconductors \cite{yampol2007controlled}. 

\section{Conclusions}

In summary, the  Boltzmann-Bloch approach is used to deliver general, non-perturbative solutions of High Harmonic Generation in semiconductor superlattices. This method allows us to investigate details of the generation processes in both spectral and time domains.  The non-approximative nature of our approach eliminates numerical errors which could cast doubt upon the origin of harmonic generation.  Thus, our study conclusively  demonstrates striking features of High Harmonic Generation when asymmetric relaxation processes are taken into account in superlattice structures. While these effects are relatively small on the odd harmonic generation, significant features appear at even harmonics  leading  to measurable effects in the GHz-THz range. Our algorithms have immediate potential to analyse the combination of asymmetric flow with parametric processes, externally applied voltages and ultrafast optical pulses. 
Future work should focus  on  investigating  thoroughly the conditions for formation of destructive electric domains in semiconductor superlattices in the case of harmonic generation due to asymmetric scattering processes.

\begin{acknowledgement}
The authors acknowledge support by the Czech Science Foundation (GA\v{C}R) through grant No. 19-03765 and the EU H2020-Europe's resilience to crises and disasters program
(H2020--grant agreement no. 832876, AQUA3S). Access to computing and storage facilities owned by parties and projects contributing to the National Grid Infrastructure MetaCentrum provided under the programme ``Projects of Large Research, Development, and Innovations Infrastructures'' (CESNET LM2015042) is greatly appreciated.

\end{acknowledgement}

\appendix
\section{Formulations of superlattice transport equations}
\label{App0}

In this section, we revisit   expressions describing a solution to SSL transport problems and having as a starting point the Boltzmann equation. The general formalism has been applied to describe transport in semiconducting devices \cite{budd1967path},  parametric amplification \cite{renk2005subterahertz}  and Bloch gain \cite{hyart2009model} in spatially homogeneous SSLs. This method allows us to deliver a general numerical solution for the influence of asymmetric relaxation effects on miniband transport model and frequency multiplication processes in superlattices in the presence of an oscillating electric field, eliminating the need for the approximative Ansatz used in Refs. \cite{pereira2017theory, pereira2017terahertz, apostolakis2019controlling, apostolakis2019devices}. The electron distribution function $f(\textbf{k},t)$ satisfies the spatially homogeneous Boltzmann equation

\begin{equation}
\dfrac{\partial f}{\partial t}= -\dfrac{\textbf{F}}{\hbar}\dfrac{\partial f}{\partial  \textbf{k}} +\int d\textbf{k}'[f(\textbf{k}')W(\textbf{k}',\textbf{k}) -f(\textbf{k})W(\textbf{k},\textbf{k}')].
\label{eq:bte3}
\end{equation}
 The second term of the right-hand side of Eq. (\ref{eq:bte3}) represents the rate  of  change of $f$ due to collisions, which is characterized conventionally by the transition probability $W(\textbf{k}',\textbf{k})d\mathbf{k}$ per unit time  that an electron will be scattered out of a state $\mathbf{k}$ into a volume element $d\mathbf{k}$  and the rate $W(\textbf{k},\textbf{k})d\mathbf{k}$ per unit tame  that an electron with wave vector $\mathbf{k}$ will scatter to a state whose vector lies between $\mathbf{k'}$ and $d\mathbf{k'}$.
 We can  rearrange Eq. (\ref{eq:bte3}) as
\begin{equation}
 \dfrac{1}{\hbar}\left( \textbf{F}\dfrac{\partial}{\partial\textbf{k}}  +\hbar \dfrac{\partial}{\partial t}+\Gamma\right)f=\textrm{\L} f
 \label{eq:bte4}
\end{equation}
Thus, if  a single and isotropic (same for all states $\textbf{k}$) relaxation rate is assumed  then $\textrm{\L} f=\Gamma f_0/\hbar$ and Eq. (\ref{eq:bte4}) has an solution in the form  $f(\textbf{k},t)=\Gamma/\hbar\int_{-\infty}^{t}dt_0 \hspace{0.1cm} f_0(\mathbf{k}(t,t_0))  
e^{-\Gamma(t-t_0)/\hbar}$. The latter equation has been  derived using the method of characteristics by Ignatov et. al. to investigate the nonlinear electromagnetic properties of SSLs \cite{ignatov1976nonlinear}.  We note that in the case of anisotropic scattering the Eq. (\ref{eq:bte4}) can be rewritten in a form of the following integral equation
\begin{multline}
f(\textbf{k},t)= \int_{-\infty}^{t}dt_0 \hspace{0.1cm} \biggl(\dfrac{\Gamma(t_0) f_0(\mathbf{k}(t,t_0))}{\hbar}  \\ + \int d\textbf{k}'f(\textbf{k}')W(\textbf{k}',\textbf{k}(t))\biggr)
 \times \exp\left\lbrace-\int_{t_0}^{t} \dfrac{\Gamma(y)}{\hbar}dy\right\rbrace 
 \label{eq:gsBoltz}, 
\end{multline}

We would like to localize the asymmetry of the electron scattering function due to interface roughness to well-defined regions of the SSL. Therefore, we assume that $W(k_z',k_z(t_0))=W_0$ if $k_z'$ and and $k_z$ both lie in within a region of the miniband for which $v_z(k_z',k_z(t_0))>0$, otherwise $W(k_z',k_z(t_0))=0$. Accordingly, the operator $\L$ is generalized into 
\
\begin{subequations}
\begin{eqnarray}
 \textrm{\L}^{+}=\frac{1}{\hbar}\left(\textbf{F}\dfrac{\partial}{\partial\textbf{k}}  +\hbar \dfrac{\partial}{\partial t}+\Gamma^+ \right) \label{eq:operator1}\\
\textrm{\L}^{-}=\frac{1}{\hbar}\left(\textbf{F}\dfrac{\partial}{\partial\textbf{k}}  +\hbar \dfrac{\partial}{\partial t}+\Gamma^- \right) \label{eq:operator2}
\end{eqnarray}
\end{subequations}
and
\begin{equation}
 \Gamma^+/\Gamma^-=1+(W_0/\Gamma^-) \int_{region^+} f(\textbf{k}')d\textbf{k}'
 \label{eq:gammaapp}
\end{equation}
\begin{figure}[t]
\centering
\includegraphics[scale=0.6]{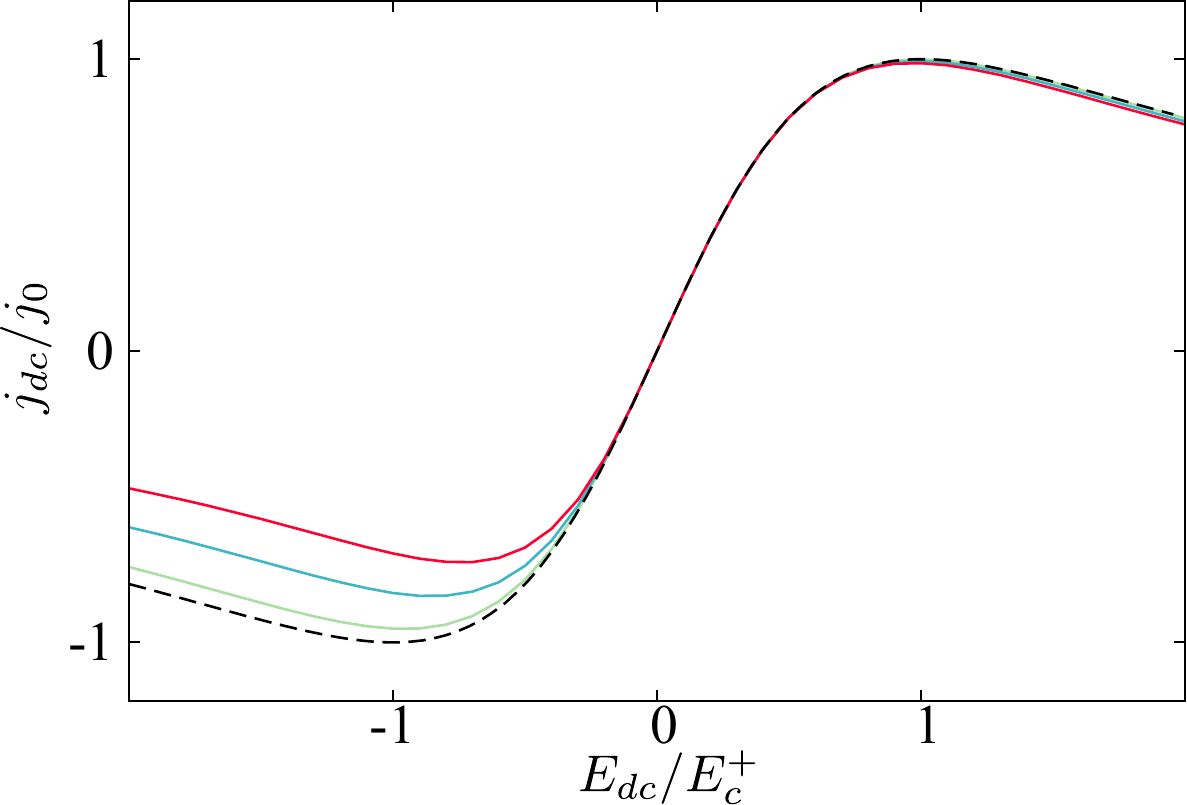}
   \caption{The current-voltage characteristics calculated for different values of the asymmetry parameter, from bottom (dashed) to top $\delta$=1,1.05,1.2, 1.4. The connection with the relaxation rates reads $\Gamma^+=eE_c^+d$ and $\Gamma^-=\Gamma^+/\delta$. The dashed curve represents the Esaki-Tsu characteristic without taking into account the time relaxation-induced asymmetries, or in other words, $\delta$=1.} 
\label{fig7}
\end{figure}
 indicating the existence of two scattering rates due to differences in interface roughness depending on the sequence of the layers. Here for simplicity we designate the region in $k$-space as $region^+$ corresponding to the high-quality interface of the SSL. Equations (\ref{eq:bte4}) and (\ref{eq:operator1}), (\ref{eq:operator2})  can be combined into the single integro-differential Eq. (\ref{eq:bte2}). The latter equation  may be solved to obtain the current  with asymmetric relaxation processes. The resulting expression  is described by 
 \begin{eqnarray}\nonumber
 j(t)=\frac{2e}{(2\pi)^3 }\int d^3k\hspace{0.1cm}{f}_0(\mathbf{k})  \hspace{0.1cm}\int_{-\infty}^{t}dt_0  \hspace{0.1cm}  \hspace{0.1cm}\dfrac{\Gamma (t_0) v(t,t_0)}{ \hbar \Delta(t,t_0)} \\ \hspace{-2 cm} \times \exp\left\lbrace-\int_{t_0}^{t}\dfrac{\Gamma(y)}{\hbar} dy \right\rbrace. \hspace{1 cm}
\label{eq:dcurrent}
\end{eqnarray}
 Note that the static current $j_{dc}$ is obtained by taking $\Delta(0,t_0)$ in Eq. (\ref{eq:dcurrent}). In that case
 the peak current $j_0=j(E_c^+)$, corresponding to the critical field $E_c^+=\Gamma^+/(ed)$ reads
\begin{equation}
  j(E_c^+)=\frac{2 d e \mid T\mid /\hbar}{ (2\pi)^3} \int_{-\pi/d}^{\pi/d} \sin (k_zd) \hspace{0.1cm} dk_z \int d^2 k f_0(\mathbf{k}).
\label{eq:jmax}
\end{equation} Figure \ref{fig7} demonstrates $j_{dc}$ versus $E_{dc}$ calculated numerically for different values of $\delta$. In contrast to the case  of $\delta=1$ (dashed curve), all other curves in Fig. \ref{fig7} exhibit maximum and minimum currents at different $\mid E_{dc} \mid$. Interestingly, as $\delta$ increases, the Esaki-Tsu peak (i.e. $j(E_{c}^+)$=$J_0$) weakens slightly whereas the peaks at the opposite bias are notably suppressed. We see  that the asymmetric current flow is dramatically enhanced  by considering scattering processes  increasingly asymmetric under forward and reverse bias. We should comment here that our approach is qualitatively different from the balance equations approach developed in \cite{ignatov1975self}  and discussed further in Refs. \cite{ignatov1991transient, alekseev1998spontaneous}. This  1D model assumed that the distribution function can be decomposed into   its symmetric $f_s=\lbrace f(\mid\mathbf{k}\mid,t)+f(-\mid\mathbf{k}\mid,t)\rbrace/2$  and anti-symmetric   $f_a=\lbrace f(\mid\mathbf{k}\mid,t)-f(-\mid\mathbf{k}\mid,t)\rbrace/2$ parts. The basic idea is that $f_s$ in the presence of  inelastic scattering processes ($\Gamma_{in}$) is allowed to relax to equilibrium distribution function $f_0$. On the other hand,   $f_a$ couples the motion only in the $z$--direction  to that in the $(-z)$--direction via  elastic scattering $(\Gamma_{el}$)  transferring the energy obtained by the electron transport along the field direction. As a result the current density-electric field dependence can be obtained by the kinetic formula $j_{dc} (E_{dc})= (\Gamma_{in} j_0/\hbar)\langle v_z(t)\exp(-\Gamma t/\hbar)\rangle $ where $\langle..\rangle$ denotes averaging over time and $\Gamma=(\Gamma_{in}\Gamma_{el}+\Gamma^2_{in})^{1/2}$. This model  predicts effectively the suppression of peak current density with the increase of $\Gamma_{el}$. It cannot, however, treat in its present form the asymmetric relaxation rates and their effects  on  harmonic generation in the presence of a time-dependent electric field, in contrast to our more general approach.
\section{Ansatz  analytical solution}
\label{App1}
 In this section we will give a recap of the analytical ansatz solution previously used  to describe asymmetric current flow and the effects of asymmetric scattering on HHG \cite{pereira2017theory,pereira2017terahertz,apostolakis2019controlling,apostolakis2019devices}. Thus, one can consider a SSL with period 
$d$ under an electric field $E_{dc} + E_{ac}\cos (2 \pi \nu t)$.  The time-dependence of the current response is then described by the Fourier basis
 \begin{align}
 j^{\nu}(t)=j^{\nu}_{dc} + \sum_{l=1}^{\infty}[j_l^{\nu,\cos} \cos (2 \pi \nu l t) +j_l^{\nu,sin} \sin (2 \pi \nu l t)],  \hspace{0.7cm} \raisetag{1.68em}\label{eq:analcur1} \\
 j_{dc}^{\nu}=\sum_{n=-\infty}^{\infty} J^2_n(\alpha) j_{dc}(U),\hspace{4.55cm} \label{eq:analcur2} \raisetag{2.5em}\\
 j_l^{\nu,\cos}=\sum_{n=-\infty}^{\infty} J_n(\alpha)[J_{l+n}(\alpha)+J_{l-n}(\alpha)]j_{dc}(U), \hspace{1.6cm} \label{eq:analcur3} \raisetag{2.5em}\\ j_l^{\nu,\sin}=\sum_{n=-\infty}^{\infty} J_n(\alpha)[J_{l+n}(\alpha)-J_{l-n}(\alpha)]K(U), \hspace{1.75cm}
\label{eq:analcur4}\raisetag{2.5em}
\end{align}
where the  dc current ($j_{dc}^{\nu}$) is given by Eq. (\ref{eq:analcur2}) and the Fourier components $\lbrace j_l^{\nu,\cos}(\alpha), j_l^{\nu,\sin}(a)\rbrace$ describe the l$^{th}$ harmonic generation. The terms $J_n(.)$ in Eqs. (\ref{eq:analcur2})--(\ref{eq:analcur4}) denote the Bessel functions of the first kind and order $n$. It can also be seen from Eq. (\ref{eq:analcur2}) that the voltage current (VI) characteristics in the presence of irradiation [$E(t)$] is given by a sum of  shifted Esaki-Tsu characteristics $j_{dc}(U)=j_0(2 U/\Gamma)/[1+(U/\Gamma)^2]$ where $j_0$ is the peak current corresponding to the critical electrical field $E_c=\hbar/(ed\tau)$. This might lead to a photon-assisted tunneling phenomenon that has been experimentally observed \cite{wacker2002semiconductor,Wacker1997}. 
Moreover, note that the term $U=eE_{dc}d+n\hbar\omega$ designates an effective potential difference instead of the plain potential drop per period due to the dc bias. The function $K(U)=2j_0/[1+(U/\Gamma)^2]$ is connected to $j_{dc}$ through Kramers-Kronig relations.  It is thus sufficient for our studies to look at the  resulting average $\mathcal{I}^2_l(\nu)=(j_l^{\nu,\cos})^2+ (j_l^{\nu,\cos})^2$ similar to Eq. (\ref{eq:PoyntVec}) ,  in order to investigate spontaneous frequency multiplication effects. The ansatz is implemented by replacing $j_0$ in Eqs. (\ref{eq:analcur1})-(\ref{eq:analcur4}) by
\begin{equation}
 j_0 = 
    \begin{cases}
     j_0& \text{$U>0$,} \\[5pt]
      j_0^-& \text{$U<0$} 
    \end{cases}, \hspace{0.2cm}
     \Gamma = 
    \begin{cases}
     \Gamma^+& \text{$U>0$,} \\[5pt]
      \Gamma^-& \text{$U<0$} .
      \label{eq:ansatz}
    \end{cases}       
\end{equation}
where the potential energy $U$ is equal to integer number of photon quanta ($n\hbar \nu$) and the asymmetry coefficient is  $\delta=j_0/j_0^-=\Gamma^+/\Gamma^-$.

\bibliographystyle{unsrt2}
\bibliography{nanophotonics2}


\end{document}